# Conductivity and Dissociation in Metallic Hydrogen: Implications for Planetary Interiors

Mohamed Zaghoo and Isaac F. Silvera
Lyman Laboratory of Physics, Harvard University, Cambridge MA 02138

**Liquid metallic hydrogen (LMH) was recently produced under static compression and high temperatures in bench-top experiments[1]. LMH is the most abundant form of condensed matter in our solar planetary structure, with Jupiter & Saturn being the largest reservoirs. Determining LMH's structure, atomic or molecular[2,3], is vital to understanding its mechanism of conduction[3-6], whereas its electronic and thermal transport properties are the key input for the magnetic dynamo action and thermal models of gas giants[7-11]. These quantities remain poorly understood. Here, we report a study of the optical reflectance of LMH in the pressure region of 1.4-1.7 Mbar and use the Drude free-electron model to determine its optical conductivity. We find metallic hydrogen's static electrical conductivity to be 11,000-15,000 S/cm, substantially higher than the only other reported experimental values[12,13], and increases with density. A substantial dissociation fraction, 65±15%, is required to best fit the energy dependence of the observed reflectance. In contrast to the interpretations of previous experiments[12], LMH at our experimental conditions is largely atomic and degenerate, not primarily molecular. We determine a plasma frequency of 20.37±2.24 eV, the highest of any known metal. The higher electrical conductivity implies that the Jovian dynamo is likely to operate out to shallower depths than previously assumed[14-16], while the inferred thermal conductivity could provide a crucial experimental constraint to heat transport models[11].**

LMH is the benchmark Coulomb system that is the simplest and the lightest of all liquid metals. Despite its apparent simplicity and fundamental significance, its thermodynamic and transport properties continue to pose outstanding challenges. Unlike other alkali metals, atomic MH is exceptional in possessing no bound electrons. Moreover, the low mass of its protonic system gives rise to a substantial zero-point energy. This LMH state is widely ubiquitous in the universe, making up about 60-70% of our solar planetary structure and the vast interiors of extra solar giant planets[15].

Over the last few decades, remarkable progress has been achieved in determining dense hydrogen's thermodynamic equation of state (**EOS**) in the region relevant to planetary conditions (see discussion elsewhere[15,17]). An important feature of the EOS is the location and the nature of hydrogen's long-predicted insulator-metal transition (IMT), where theoretical predictions differ by more than factors of four in both pressure, P, and temperature, T[17]. This IMT phase boundary was recently mapped in static experiments where both P and T were measured[1]. However, the transport properties of LMH remain less-well understood[18,19]. A key long-standing issue that pertains to these properties is the mechanism of electronic conduction in metallic hydrogen: thermal excitation of carriers across a reduced mobility gap[4,12,13] versus conduction by free electrons in a degenerate atomic system[2,3]. The central question is whether metallization at these conditions occurs first in the molecular phase or if it proceeds by dissociation into a conducting atomic phase, as predicted by most theories[17]. Accurate determination of the optical conductivity from the reflectance, as well as its density dependence, are experimental probes that can distinguish between these two models of metallization[4].

Theoretically, the confluence of quantum and thermal effects, as well as the strong coupling between the electrons and the protons, make it difficult to accurately calculate electrical and thermal conductivity. Theoretical models have to rely on different assumptions about the ionic system structure[2,7,20], the degree of ionization[19-21], and proton-proton correlations[3], all of which remain challenging to adequately describe in these models. Ab-initio density functional theory (DFT) simulations differ substantially in their calculated conductivities, depending on the functionals used[18,22] (see also discussions in the Supplementary Materials of ref[23]). Experimentally, pioneering shock-wave experiments have reported plateauing of electrical conductivity around 2000 S/cm in the region 1.4-1.8 Mbar[12]. It was argued that the plateau of conductivity was due to thermal smearing of the band gap and that metallization occurs in the molecular phase with 5-10% dissociation fraction[4,12]. We note that the reported values of conductivity are a factor of 3 less than the Mott-Ioffe-Regel (MIR) minimum metallic conductivity criterion, 6000 S/cm[13,24], and a factor of 5 to 10 lower than values predicted with different theoretical models for fully ionized metallic liquids[2-4,6,7,19].

Here we report the first measurements of the optical conductivity of bulk metallic hydrogen at planetary interior conditions in the pressure region of 1.4-1.7 Mbar and measured temperature of 1800-2700 K, which is comparable to the conditions studied by Weir et al.[12,13]. Pressurized hydrogen was pulse-laser heated, and time resolved spectroscopy was used to measure the optical reflectance as a function of T, using a different experimental setup from that in ref[1] (see Fig.S1). The duration of the laser pulse, 290 ns, is sufficiently long to achieve local thermal equilibrium yet short enough to inhibit sample diffusion into metallic gaskets and diamonds (see SI).

Optical reflectance of hot dense hydrogen samples was measured simultaneously or separately at three wavelengths: 514 nm, 633 nm, and 980 nm. We observed an abrupt increase in reflectance above a certain transition temperature consistent with previously results[1] (see Fig.1). At threshold, the LMH film is thin and semi-transparent. As the laser power is increased the film heats and thickens until the transmission is essentially zero and reflection corresponds to that of a bulk metal. This value of reflectance is consistent with values obtained in previous shockwave and static experiments[1,23,25,26]. Once the reflectance saturates, it is weakly dependent on temperature, as expected in a degenerate metallic system. We further contrast our observed reflectance to that expected for hydrogen if it was semiconducting with a finite gap and the carriers are thermally activated, as previously suggested[12] (see Fig.1), finding strong disagreement with our current measurements.

Reflectance measurements are amenable to Fresnel analysis. The measured reflectance at a given frequency is $R(\omega)=|(N_D-N_H)/(N_D+N_H)|^2$ where $N_D$ is the index of refraction of molecular hydrogen/diamond layer and $N_H$ is the complex index of refraction of LMH (see SI). Both the observed saturation of optical reflectance, degeneracy, and the distinct wavelength dependence of that conduction are important conditions for application of the free electron model. We have thus analyzed the optical response of LMH, only in the degenerate limit, using the Drude free electron model, which has previously been extensively used for LMH[2,5,6]. This restriction is

further justified below. In this model, $N_H^2 = 1 - \omega_p^2/(\omega^2 + j\omega/\tau)$ where $\omega_p$ is the plasma frequency, which is directly related to the carrier density $n$ by $\omega_p^2 = 4\pi n e^2/m_e$, and $\tau$ is the relaxation time. We define a dissociation/ionization fraction Z as the ratio of the carrier to ion density. Thus, if all molecules are dissociated, Z=1, while if all molecules are ionized or half atomic-molecular, Z=1/2. Accurate measurements of R(ω) can thus be used to determine τ and ω_p from which the carrier density or the degree of dissociation can be deduced. We performed a least-squares non-linear fit of our measured R(ω) to extract these parameters (see Fig. 2).

At 140 GPa, our reflectance data is best fit to ω_p =20.4±1.56 eV and τ of 1.3±0.2 x10$^{-16}$ s, yielding a dissociation fraction Z of 0.65±0.15. We determine σ_dc= 11,000±1100 S/cm, a factor of 2 higher than Mott's minimum metallic conductivity at this pressure. At 170 GPa, our new data, as well as those reported in ref[1] are best fit to ω_p =21.76±2.8 eV and τ =1.6±0.3 x10$^{-16}$ s. This corresponds to σ_dc =15500±1500 S/cm. We contrast our optical reflectance, and the determined Drude fits, to that expected for LMH using the minimum relaxation times, $\tau_{min}$ prescribed by the MIR limit. Such a limit, which is often employed by shock-wave experiments, cannot account for the observed reflectance of LMH. Moreover, at the onset of the insulator-metal transition boundary, where reflectance is still rapidly increasing, our optical data <u>cannot</u> be fit to the Drude model, suggesting a non-free electron like density of states (see Fig 2 top left panel). As the dissociation fraction increases, LMH becomes more Drude–like and when the carriers are degenerate, the free electron character is established. This result provides a further justification for the validity of the Drude model in the degeneracy region. The discussed crossover in behavior has been demonstrated theoretically[18]. The real part of the optical conductivity is plotted in Fig.2 where the relaxation time, determined at 1.26-2.33 eV, is assumed to be the same as that at zero energy. We have investigated a possible deviation from this assumption through the empirical Smith-Drude model[27]. In this model an additional backscattering term is introduced that shifts the Drude Lorentzian peak to a non-zero frequency. We show that a transfer of the oscillator strength to the infrared (below our measured 1.2 eV) will only have a minor effect on the DC conductivity (see Fig. 2 top left panel).

We have calculated the electronic contribution to thermal conductivity of LMH using the Wiedemann-Franz law, which is well substantiated by theory[19]. In Fig. 3, we compare our electrical and thermal conductivity values to several theoretical predictions using various models of different degrees of sophistication. Our LMH DC conductivities and relaxation times are in very good agreement with DFT-MD[2,6] and CEIMC[5] calculations that predict σ_dc=13000 S/cm and τ= 3.0± 0.1x10$^{-16}$s and σ_dc=8600 S/cm and τ= 3.1± 0.3x10$^{-16}$ s, respectively, as well as those calculated for fully ionized hydrogen plasmas, albeit at higher T[21].

Our electrical conductivities are a factor of 6-8 higher than those previously reported by Weir et al.[12]. The uncertainties in those measurements ranged from 25-50%[12]. The origin of discrepancy is unknown, with temperature being unlikely factor, since our optical data shows a clear saturation as a function of increasing T. Moreover, assuming that conductivity may suffer a reduction at increasing T, its limiting value (MIR limit) is still 3 times higher than reported in ref.[12]. Possible sources of uncertainties for those measurements could include inaccuracies in estimating the shocked cryogenic cell thickness at the highest compression, the value that relates the measured resistance to resistivity (see SI). We note that our reflectance measurements in the bulk limit are insensitive to this consideration. Experiments on LMH should, in principle, observe increasing reflectance as a function of density, provided the temperatures are relatively low.

Within the experimental errors, the energy dependency of the optical reflectances is well described by the Drude model. A large dissociation fraction within the free electron model is required to best fit that energy dependency. We note that irrespective of our choice of $\tau$, a 5-10% dissociation fraction inferred from ref[12,13] cannot fit our reflectance data (see Fig.S5). Our observed increasing optical conductivity as a function of pressure is also consistent with this conclusion, where a rising density of electrons as a result of dissociation or ionization leads to a further increase in conductivity. A negative slope of the metallization phase line revealed from our data is also consistent with the dissociation model; the dissociation energy decreases with density and thus the transition temperature decreases. In contrast to the interpretation proposed in previous experiments[12,13,23], LMH at the conditions of our experiments is largely atomic and degenerate. The transition is also abrupt, consistent with a first-order phase transition, with no observable concomitant interband transitions in the frequency range investigated. Our results, thus, do not support a conduction mechanism by thermal smearing of a reduced Mott-Hubbard gap (see Fig. 1). A mechanism of band overlap in the molecular phase has also been suggested where one electron per molecule contributes to conduction or the equivalent of 50% dissociation fraction[4]. Our current measurements cannot distinguish between a mixed atomic-molecular metal and a fully molecular metal within the uncertainties of the fit (see SI). Additional low frequency reflectance measurements could resolve this uncertainty.

The conductivity of LMH reported in this letter is representative of that at P-T conditions of ~ 0.84 of Jupiter's radius, $R_J$.[28] It corresponds to a magnetic diffusivity of ~ 0.56 x $10^4$ $cm^2$/s, a factor of 8 lower than previously inferred from Weir et al.[12] and now roughly similar to new estimates of Earth's iron core (see SI). Our values are also considerably higher than the ab-initio transport values calculated along the Jovian adiabat at similar pressures[29], where the utilized DFT functionals evidently underestimate conduction. The implication is that the dynamo is likely to operate out to a pressure that is much lower than that at $R_J$ =0.84, into a regime where hydrogen is not fully ionized. If metallization along the Jupiter adiabat is continuous, as both theory and previous experiments suggest, then the depth at which hydrogen conductivity reaches values corresponding to a magnetic diffusivity sufficient to sustain the dynamo (~$10^7$ $cm^2$/s) may extend to more than $R_J$=0.91. This is much lower than most previous estimates[8,14,16] and even likely lower than the value inferred by recent dynamo simulations[10,29] based on the ab-initio transport results. Furthermore, the dissociation-based picture argued in this letter means that LMH transport coefficients should scale with density or the Jovian radius, consistent with our observation for optical conductivity. This is especially significant for most anelastic simulations of the Jovian dynamo that otherwise prescribe a characteristic constant conductivity for depths below the metallic transition region[16].

Our inferred thermal conductivity values are considerably lower, a factor of 4-5, than the theoretical estimates in current use for Jovian planets heat transport models[7,30]. Revised estimates should thus provide a crucial experimental benchmark for future models, particularly those proposing layered convection as a likely origin for Saturn's strikingly higher luminosity[9] or the anomalously large radius of several transiting hot Jupiters[11].

**Acknowledgments**
We thank Ashkan Salamat, Rachel Husband, Ranga Dias, and Ori Noked, as well as Bill Nellis for discussions. The NSF, grant DMR-1308641, the DOE Stockpile Stewardship Academic Alliance Program, grant DE-NA0003346, and NASA Earth and Space Science Fellowship Program, Award NNX14AP17H supported this research. Preparation of diamond surfaces was performed in part at the Center for Nanoscale Systems (CNS), a member of the National Nanotechnology Infrastructure Network (NNIN), which is supported by the National Science Foundation under NSF award no. ECS-0335765. CNS is part of Harvard University.

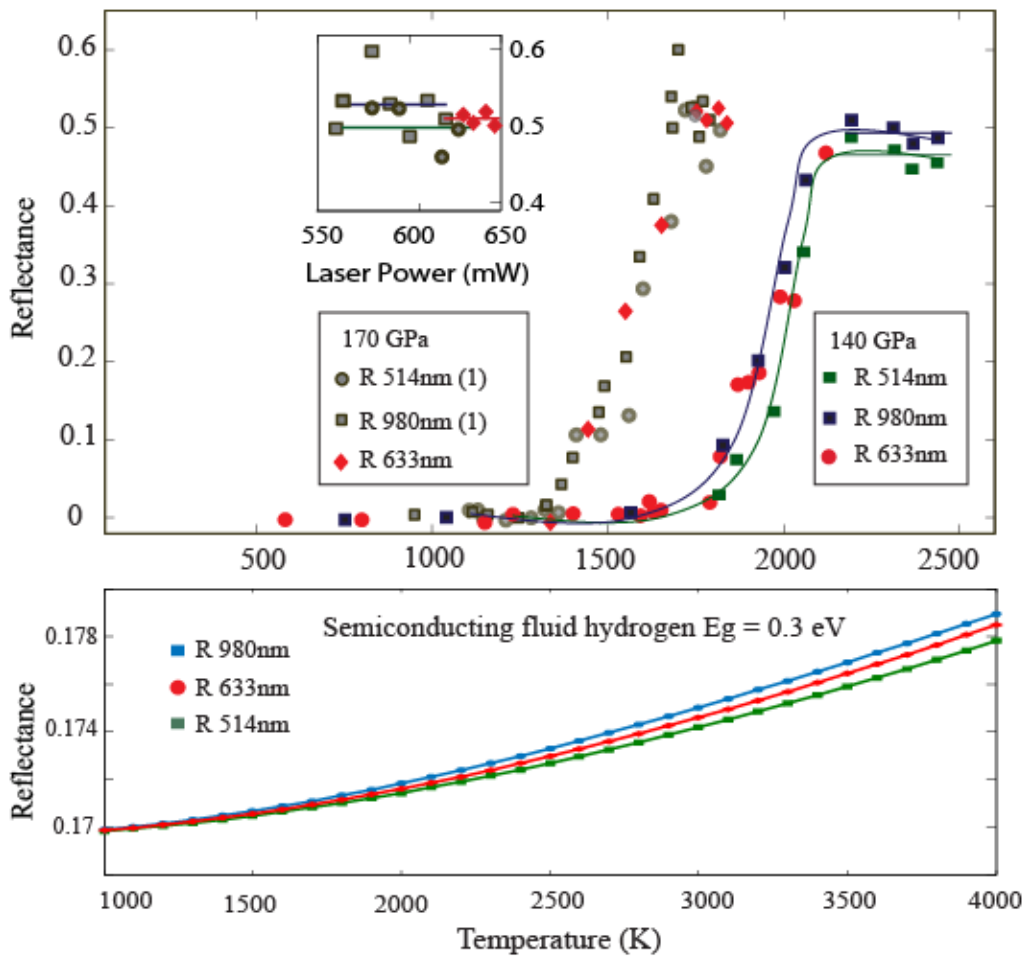

Figure 1

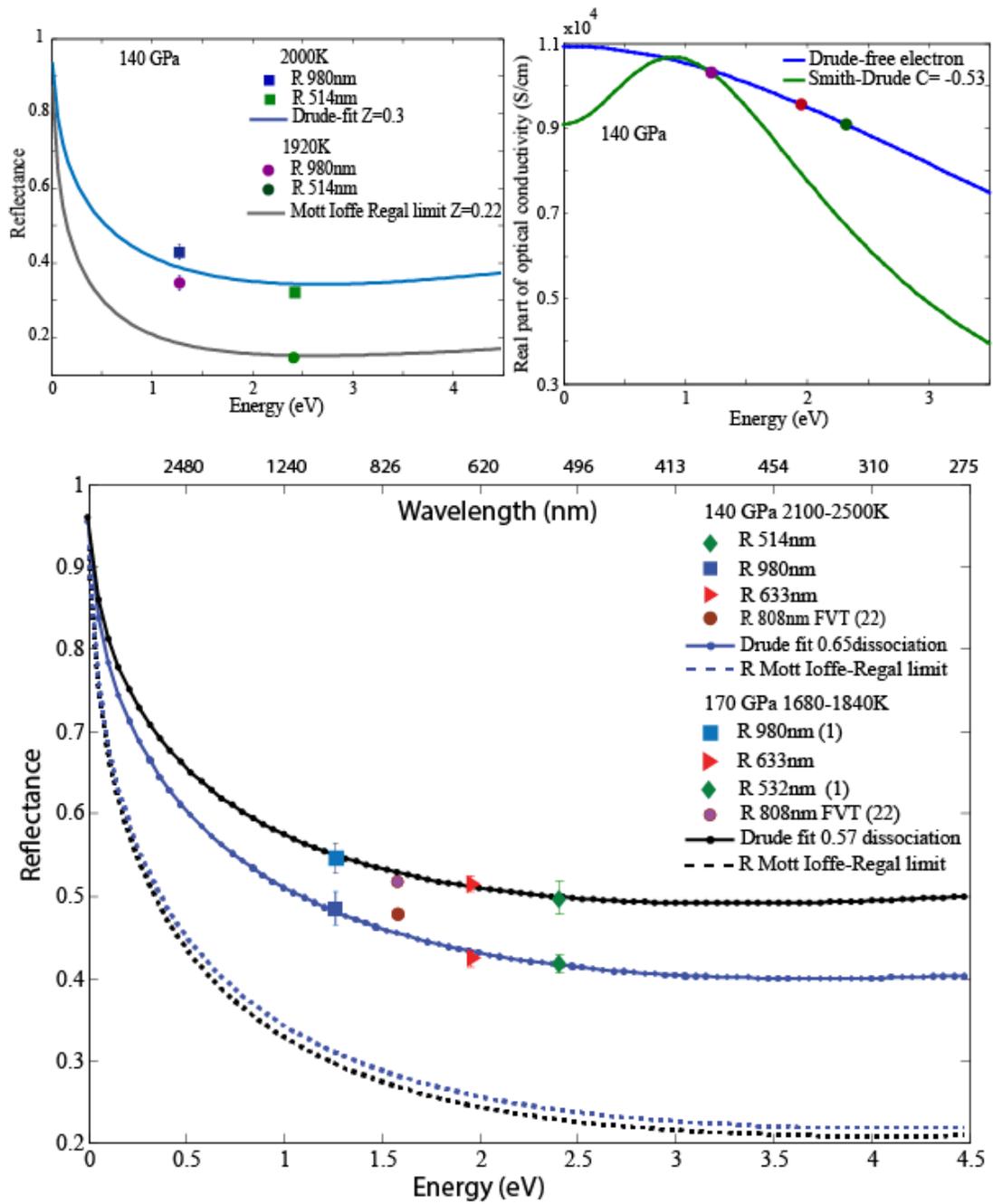

Figure 2

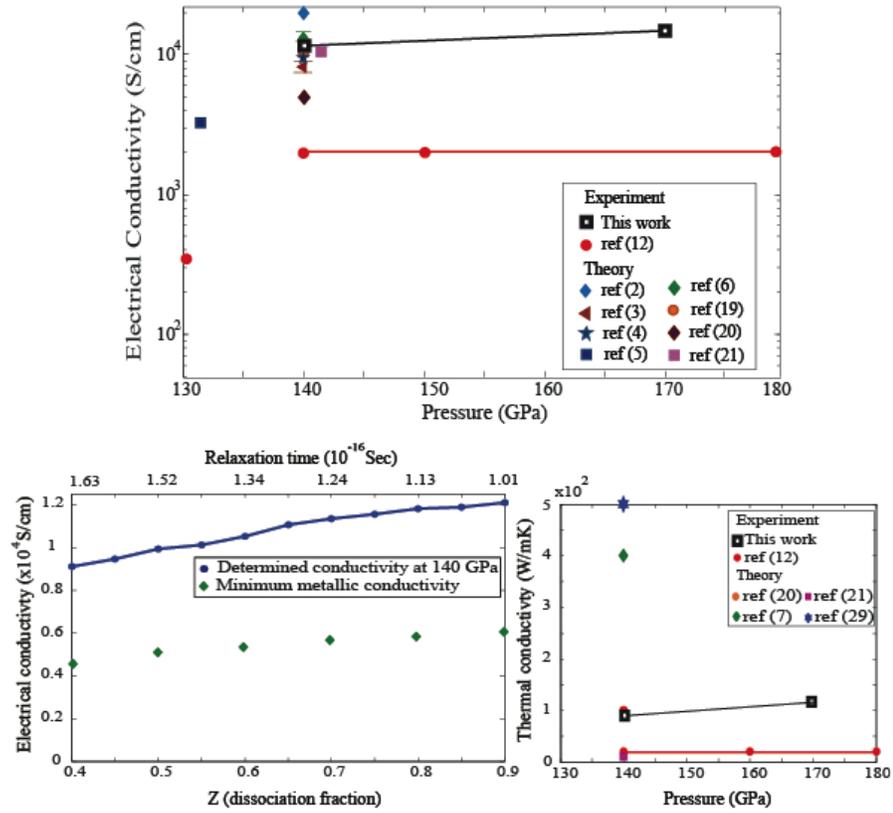

Figure 3

## Figure Captions

**Fig.1.** Dense hydrogen reflectance as a function of temperature plotted for three wavelengths, 514, 633 and 980 nm. Above a certain transition temperature, liquid hydrogen metallizes and reflects incident probe light. Lines are guides to the eye. The relatively larger scatter in the 170 GPa LMH bulk reflectance data, shown in detail in the inset, is because each wavelength measurement was collected separately, while for 140 GPa reflectances at 514 and 980 nm were measured simultaneously. Bottom panel shows calculation of expected reflectance due to thermally activated carriers plotted against temperature if hydrogen were semiconducting with a 0.3 eV band gap (see SI).

**Fig.2.** Bottom: Bulk liquid metallic hydrogen reflectance as a function of energy or wavelength is plotted for 140 (blue line) and 170 GPa (black line) pressures in the degeneracy region (saturation of reflectance). Optical conductivities are higher for increasing pressures as expected in dissociation-induced metallization. The Drude fit to the experimental data is derived from a least-squares fit to the energy dependence of measured reflectance. The dotted lines are the calculated reflectance using the MIR limit. Top Left: LMH reflectance below the degeneracy region at 140 GPa at two temperatures. At 1920K, with low reflectance (see Fig. 2), the wavelength dependence cannot be fit with the Drude model, however at higher temperatures and reflectance, as the dissociation increases close to degeneracy, the free electron character is established, the data can be fit. Top Right: Real part of the optical conductivity plotted as a function of energy for two different models: the Drude-free electron versus the empirical smith-Drude[27] model with a backscattering term. Solid circles represent the measured values for optical conductivity. We show that a non-Drude like behavior below 1.2 eV will only have a small effect in the determined conductivity (details are in the SI).

**Fig.3.** Top: DC electrical conductivity of liquid metallic hydrogen as a function of pressure on a logarithmic scale, compared to several theoretical predictions. Uncertainty is about 10-12% in the experimental data. Lines are guides to the eye.
Bottom Right: Thermal conductivity of LMH as a function of pressure at 4000K determined from the Wiedemann-Franz law[19], compared to several theoretical predictions. Bottom Left: The electrical conductivity of LMH as a function of dissociation fraction and relaxation times determined from the fits to our measured data. The green diamonds are the minimum metallic conductivity criterion calculated as a function of density using Mott's formula $\sigma_{min} = e^2/3\hbar a$[24]. Lines are guide to the eye.

## Supplementary Information

Hydrogen was pressurized in a Diamond Anvil Cell (DAC) at room temperature using rhenium gaskets. Diamonds were coated with a 50 nm layer of alumina that acts as a diffusion barrier against hydrogen. Samples at static loads (fixed density) were pulse-laser heated using an Nd-YAG infrared laser (20 KHz repetition rate and 290 ns pulse width) to peak temperatures as high as ~2700 K. Since molecular hydrogen is transparent and will not absorb the laser energy, a thin (~7 nm) tungsten film that performs as a laser absorber was deposited on the alumina layer of one of the diamonds (and covered with a 5 nm protective layer of alumina). This was heated by the laser which conductively heats the hydrogen sample pressed against its surface[1,31]. The sample region of our DAC is shown in Fig.S1, as well as the reflectance geometry. The hydrogen layer adjacent to the W film is in local thermal equilibrium with the W during this relatively long pulse, as electrons and translational motions thermalize within ~10 ps.[32] The energy/pulse is relatively small (~0.005 mJ), so that the diamonds remain at room temperature, on average. Pressure is determined by the shift of the Raman active vibron of the hydrogen. The pressure was measured at room temperature before and after the heating run to ensure the integrity of the sample. Temperature was measured by pyrometry techniques. At high pressures, the molecular hydrogen is a few microns thick, while we estimate that the heated part of the sample that metallizes in the bulk limit is of submicron thickness. The details of the experimental setup, technique and optical setup are described elsewhere.[1]

**Reflectance analysis**

Time-resolved reflectance measurements were conducted at fixed wavelengths, 514, 633, and 980 nm, using CW solid-state lasers. The reflected light off of the W/hydrogen interface was monitored as a function of temperature and pressure on Si detectors with 2ns rise time. At low temperatures, the W layer dominates reflectance when hydrogen is still in the molecular phase. Above a certain transition temperature, the hydrogen sample metallizes and starts reflecting. LMH reflectance is defined as reflectance signal above that temperature normalized to that below the transition to LMH. The phase line delineating this boundary between the insulating and metallic phase, as well the thermodynamic path for the current experiments, are shown in Fig.S2. As more energy is delivered to the LMH sample, its thickness grows and temperature increases. Once the LMH layer is thick enough, $R(\omega)$ plateaus at a fixed bulk value that is consistent with previous shock-wave experiments[23,25,26]. The mean of $R(\omega)$ at that saturation limit is the experimentally reported value used for the Drude fit and the standard deviation determines the error bar. In the bulk limit, the interface between the metallic and molecular hydrogen is much smaller than the wavelength of the probe light and thus it could be approximated as a specularly smooth surface. At 140 GPa, the index of refraction of solid molecular hydrogen, and to less than a percent fluid molecular hydrogen, is ~2.4[33]. This is close to that of diamond and so we treat the molecular hydrogen diamond as one interface (FigS1). We note that the choice of this pressure region to investigate metallic hydrogen bulk reflectance properties is thus convenient, as we do not have to deal with multi-layer interfaces.

Eight optical runs, monitoring $R(\omega)$ as a function of T, were conducted on two different samples, where $R(\omega)$ values of bulk LMH showed excellent reproducibility (See Fig.S3). Some systematic uncertainties due to roughening of the W/hydrogen interface with time occurred, but should be rather small. The $R(\omega)$ data at 140 GPa were collected simultaneously at 980, 514 nm, with an additional point at 633 nm, on two samples and had better statistics than the 170 GPa data. Measurements of $R(\omega)$ at 170 GPa were conducted separately on all wavelengths, but on the same sample. Additional measurements at lower or higher frequency could further constrain the fits, but will not change the overall results. Figure S4 shows the weighted least-squares-fit at 140 GPa. The plasma frequency (carrier density) and the relaxation time were extracted from the fits using the Drude reflectance formula described in the main text.

Although the best fit to our three wavelength reflectance results yields a dissociation fraction, Z = 0.65± 0.15 with an R-square value of 0.95. We investigated the range of fitting parameters by imposing restrictions. For example, if we restrict the value of Z to 0.4 to 0.5, we find an R-square value of 0.70 to 0.90; and $\tau=1.54\pm0.17\times10^{-16}$ s. At 170 GPa and 1700K, our reflectance data is best fit to a $\omega_p$ =21.76±2.8 eV and $\tau=1.6\pm0.3 \times10^{-16}$ s, yielding a dissociation fraction Z=0.57±0.14. Within the uncertainty in the fits, this dissociation fraction is comparable to that at 140 GPa at 2500 K. However, we note that the difference in temperature range where the optical data were fit could explain why it is not higher at increasing pressures. Since the static Drude electrical conductivity depends linearly on both the dissociation fraction (plasma frequency) and relaxation time, the conductivity is weakly dependent on Z (see bottom left panel, Fig.3). For Z < 0.35; R-square becomes negative.

In Fig. S5, we show that a dissociation fraction of 5-10% of the carrier density, inferred from shock experiments by Weir et al at similar densities and temperatures cannot fit the energy dependency of the measured reflectance. A substantially higher dissociation fraction is required to explain the high reflectance.

**Smith-Drude Model**
A common assumption in optical conductivity experiments is that the optical constants determined at the measured frequencies (1.2-2.33 eV in our case) are the same as those at low or zero frequency. In the Drude free-electron model, this assumption means that the determined relaxation time τ (1.2-2.33 eV, where the data is very well described by the Drude model) is equal to τ at ℏω=0 eV. However, to study possible deviation from this scenario at energies lower than 1.2 eV, we have analyzed our data using the empirical Smith-Drude (SD) model for conduction in liquid, disordered, or poor metals[27]. The SD model was initially proposed to account for the anomalous optical conductivity of liquid mercury in the mid- IR, and has since been applied to poor/dirty metals. It introduces a backscattering term, c, to the Drude-generalized formalism, which shifts the oscillator strength away from zero frequency to higher frequencies and depresses the DC static conductivity. The generalized SD form is, to first order

$\sigma(\omega) = \frac{ne^2\tau}{m(1-i\omega\tau)}[1 + \frac{c}{1-i\omega\tau}]$.

The model contains three unknowns, $\tau, n, c$. We would like to investigate if a fit of our data is possible that reproduces our measured wavelength dependence of the optical conduction over the range of (1.2-2.33 eV), and yet have a DC conductivity that differs from the Drude model. We found that no possible values of $\tau, n, c$ could do this. We have thus relaxed that restriction and required that the putative shifted peak due to backscattering is lower than 1.2 eV (our lowest measured energy) and it matches the measured value for optical conduction, i.e.

$\sigma_{Drude}(1.2\ eV) = \sigma_{Smith-Drude}(1.2\ eV)$

A limited range of values for c, Z and $\tau$ were found that could satisfy these two conditions. In Fig. S6, we plot the real part of the optical conductivity of the Smith-Drude model using this expression

$$\sigma_{Smith-Drude}(\omega) = \frac{ne^2\tau}{m(1+\omega^2\tau^2)}[1+\frac{c(1-\omega^2\tau^2)}{(1+\omega^2\tau^2)}]$$

for c=-0.53, $\tau = 4.53x10^{-16}s$ and Z=0.334 in comparison to the Drude free-electron model. Higher values of Z, up to 0.41, could exist for slightly lower c, and $\tau$ as low as $3.8x10^{-16}$s. The two most important results are, first: even within the SD model, a very high dissociation fraction is needed to account for the observed reflectance in the infrared. Second, the DC conductivity is only lower by a 15-20% from our quoted values in the Drude-free electron model. As c goes to zero, the Drude-free electron behavior is recovered. We conclude that a deviation from the Drude behavior at lower frequencies is unlikely to change our main conclusions based on the Drude model.

**Thermal excitation in semiconducting hydrogen**

Previous shockwave experiments have suggested that dense fluid hydrogen at P-T conditions similar to that of our experiment is a semiconductor with a finite gap, $E_g$ of 0.3-0.4 eV. Additionally, it was argued that conduction is caused by thermal excitation of the localized carriers across that gap. We have thus calculated the reflectance signal for bulk hydrogen as a function of T if it were intrinsically semiconducting, to contrast it with our observed reflectance measurements (Fig. 3). In semiconductors, thermal excitation of carriers from the valence band to the bottom of the conduction band depends both on the density of available states in the conduction band and the probability of occupation as described by the Fermi-Dirac integral, given by

$f_m[x] = 2/\sqrt{\pi}\int_0^\infty y^m/(1+e^{y-x})dy$. The free-carrier concentration in the conduction band is $n_i = N_S f_{1/2}[-E_g/2K_BT]$, where Ns is the number per unit volume of effectively available states, given by $2(\frac{m_{eff}K_bT}{2\pi\hbar^2})^{3/2}$. Once $n_i$ is known, the index of refraction and Fresnel reflectance as a function of T can be calculated using the same formula provided in the main text. We have assumed that the gap doesn't change significantly over a temperature range of 0.3 eV, as was also inferred in previous shock analysis. We used a relaxation time similar to that determined from our data to make the comparison simpler. However, the overall qualitative behavior (R values and dependence on T) is rather weakly dependent on either choice. The result is shown in Fig. 1.

**Minimum metallic conductivity and mean free path (mfp)**

In the Ioffe-Regal limit[34], the mean-free-path, the average distance electrons travel between collisions, cannot be shorter than the interatomic distance, a. This limit defines the Mott-Ioffe-Regal minimum metallic conductivity[24] and is given by $\sigma_{min} = e^2/3\hbar a$, where ℏ is the reduced Planck's constant, and a ~ $(n_i)^{-1/3}$ is the interparticle spacing with $n_i$ being the ion density.

For metallization in the molecular phase, initially suggested by Weir et al.[12], one electron per molecular ion contributes to conduction, so at the metallization pressure of 140 GPa, the density from ref[35] is 0.385 mole/cm$^3$, corresponding to a molecular number density of 2.318x10$^{23}$ while a = 1.628 Å. In this scenario, $\sigma_{min}$~ 4984 S/cm. If metallization occurs in the atomic phase, where the molecules are fully dissociated, then the carrier density is twice as high as in the molecular phase, 4.6x10$^{23}$ mole/cm$^3$, and $\sigma_{min}$= 6280 S/cm. Our conductivity results are consistent with metallization in the higher carrier density atomic phase. The mfp is generally assumed to be vf x $\tau$ where vf is the Fermi velocity and $\tau$ is the relaxation time. At the metallization density of 0.385 mole/cm$^3$, $\tau$ is determined from reflectance data to be ~1.2x10$^{-16}$ and the mfp is about 3.317 Å compared to an interatomic spacing, a~ 1.29 Å.

**Uncertainties in the dynamic shockwave conductivity experiments**

In the four-probe arrangement used in the 1996 shock-wave conductivity experiments; a calibration calculation is used to determine the cryogenic sample cell dimension at different compressions[12,13]. This calibration calculation is used to relate the measured resistance value to the reported resistivity at these compressions. These calculations assume an equilibrium current density over the cross sectional area of the two current measuring electrodes, i.e. that the electrodes are flush at the sapphire anvil/hydrogen interface. However, as noted in Nellis et al.[13], the electrodes intrude into the hydrogen samples "a few 10 μm" because of different shock impedances of steel and sapphire. This is comparable to the sample thickness of 50 um, and can perturb the current density leading to additional uncertainty in the conductivity. This could possibly explain the difference between our determined conductivities and those earlier reported by Weir et al.[12]

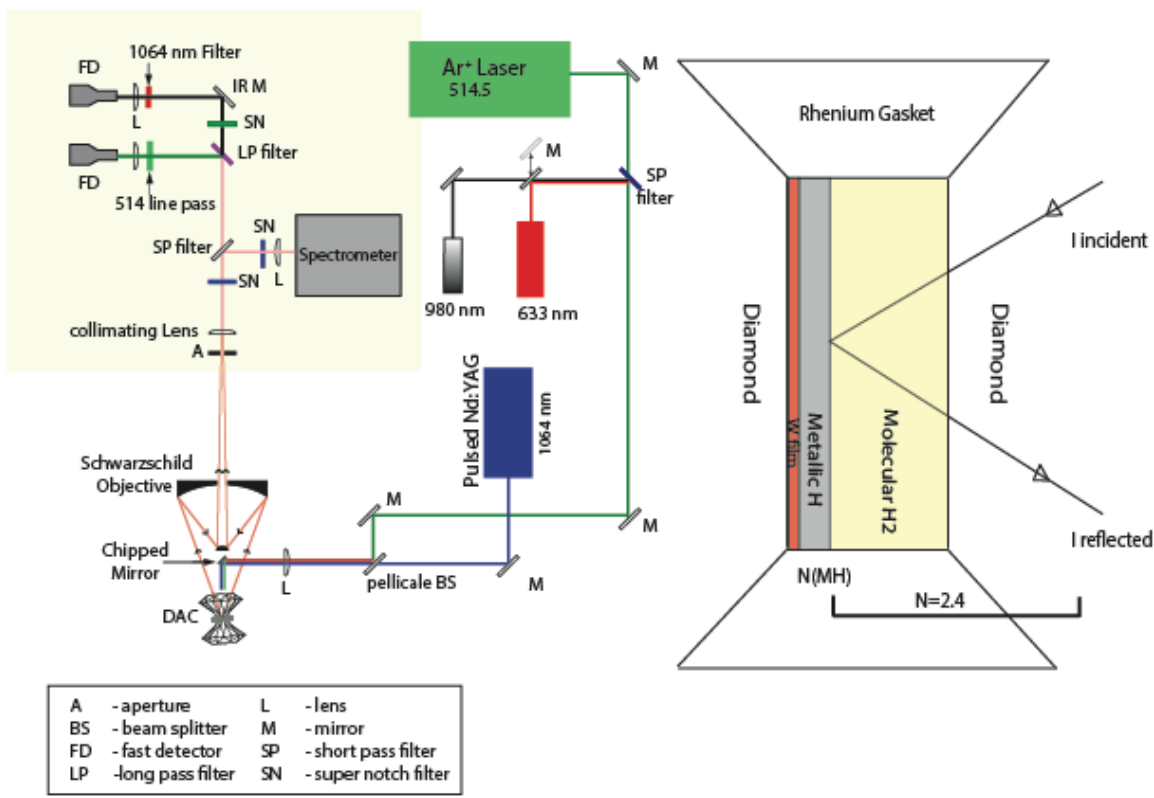

**Fig. S1.** Left: The experimental layout used in the current experiments to simultaneously measure optical reflectance at two different laser frequencies as a function of T. In our previous experiments[1], only one-laser light measurement was possible at a time, rendering accurate determination of optical constants difficult. Right: a schematic of the DAC interior showing the reflectance measurements arrangement. The interfaces between the layers are to a very good extent specularly smooth. At the experimental conditions, the diamond and the molecular hydrogen layer have approximately the same index of refraction. Estimation on the thickness of each layer is provided in the text.

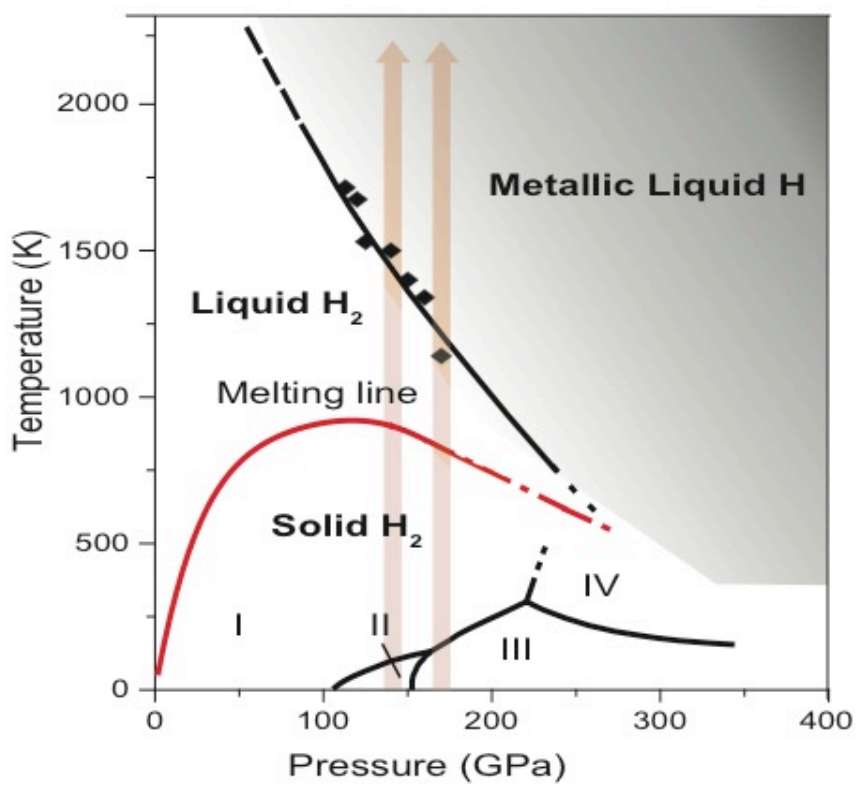

**Fig. S2.** The phase diagram of hydrogen showing the insulator-metal transition boundary and the data points from ref.[1]. The continuous curves are fitted to experimental data; the broken lines are extrapolations. The two thick arrows show the thermodynamic pathway used for the current measurements.

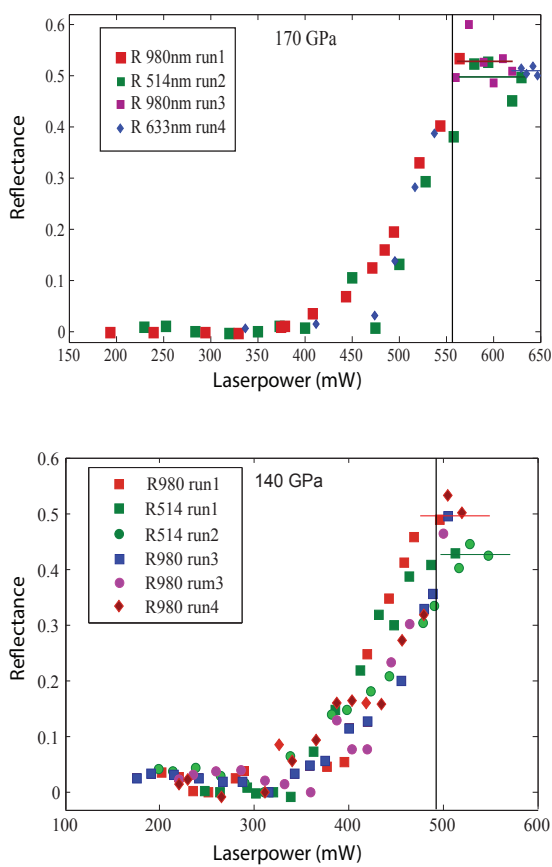

**Fig. S3.** Raw reflectance data as a function of laser power collected at eight runs for different wavelengths. The vertical black solid line denotes the region where the metallic hydrogen layer reaches the bulk limit at which reflectances saturate. The horizontal lines are guides to the eye for the average value of this reflectance saturation for respective wavelengths.

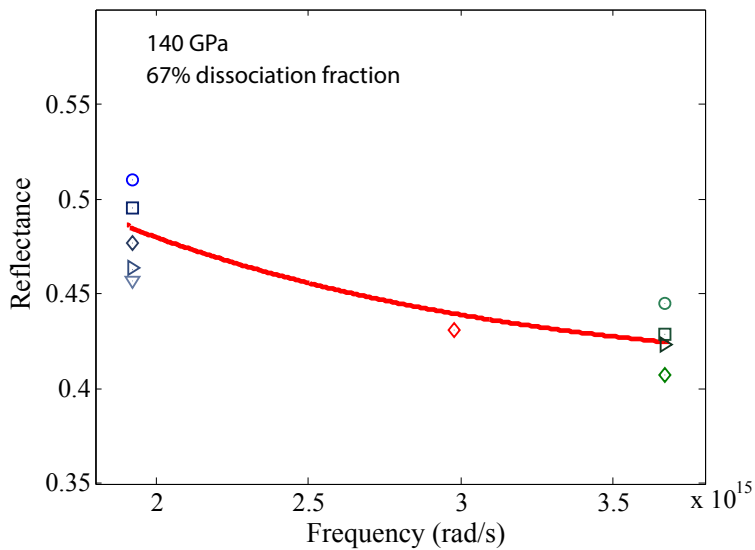

**Fig. S4.** LMH reflectance data at 140 GPa as a function of angular frequency fitted to the Drude reflectance formula to extract the Drude parameters, plasma frequency, and relaxation time. The data used for the fit are the reflectance points shown in Fig. S3 to the right of the vertical line, corresponding to reflectance saturation.

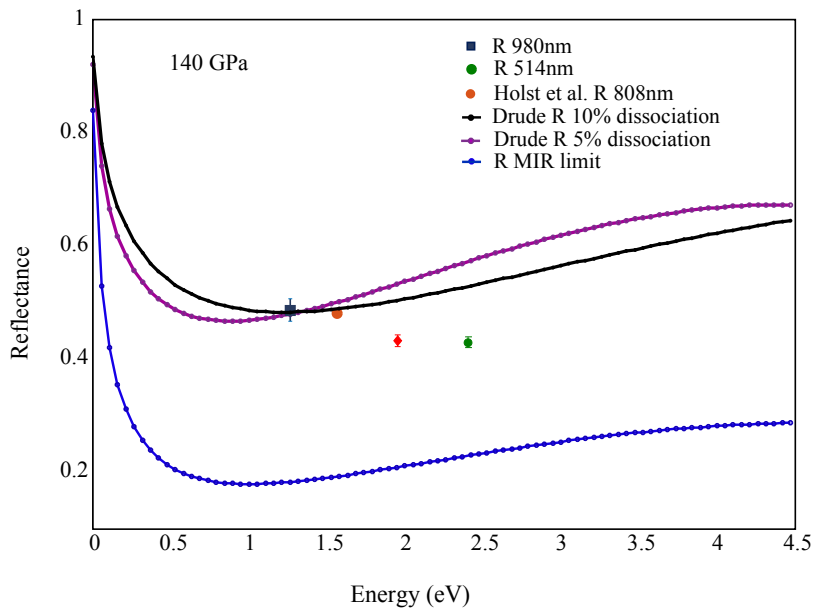

**Fig. S5.** LMH reflectance data as a function of energy at the 140 GPa. Two different Drude reflectances are calculated for 5%, and 10% dissociation fraction reported in ref[12]. We show that irrespective of our choice of $\tau$, a small dissociation fraction cannot fit the energy dependency of our reflectance data. The calculated reflectance using the Mott-Ioffe-Regal limit is plotted for comparison.

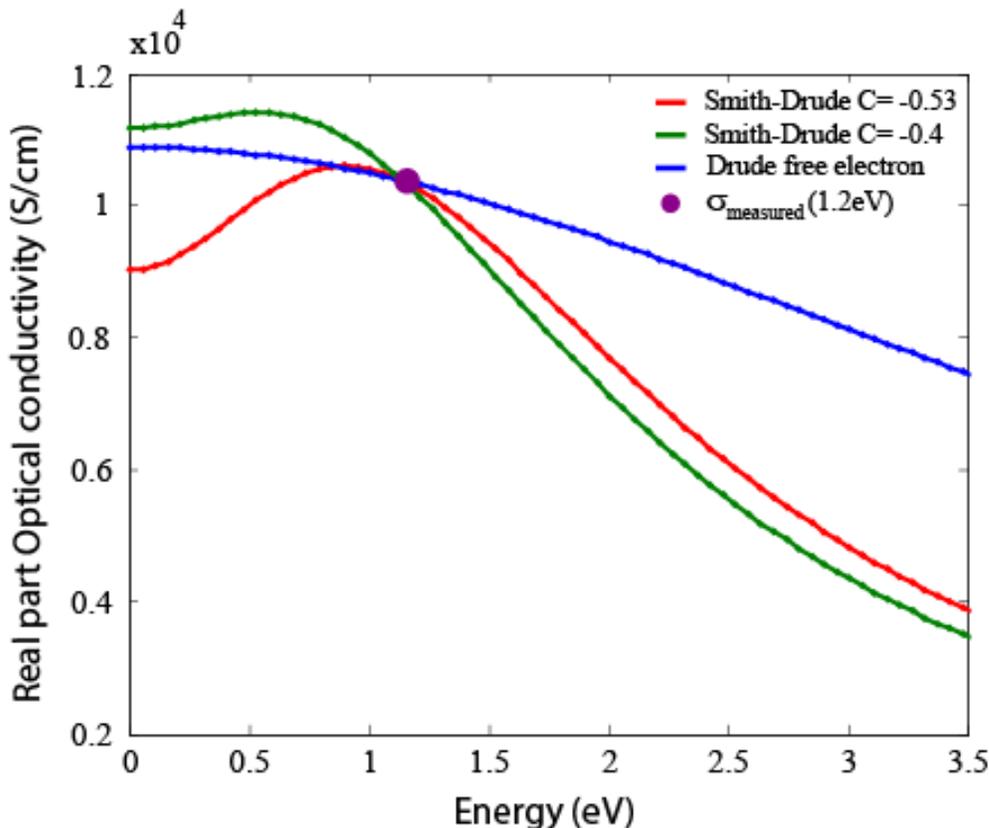

**Fig. S6** Real part of the dynamic conductivity plotted as a function of energy using the Smith-Drude model for different c values compared to the Drude-free electron model. Notice that as c increases, the infrared peak shifts to lower frequencies and the optical conduction becomes more Drude-like. The purple point is the optical conductivity datum from the reflectance measurement at 1.2 eV.

**The Free electron Model**
The Drude free-electron model is a fundamental approach to studying equilibrium transport properties of conducting systems. Indeed it has been successfully employed to describe reflectance measurements (optical conduction) in shocked compressed water[36], helium[37], fused silica[38], diamond[39], LiF[40], $Al_2O_3$[40] and hydrogen[5,6,25,26,41] at more extreme conditions, up to 20 Mbar and 20000 K. Unlike all of these previous cited shock-wave experiments that relied on a single-wavelength reflectance measurements and simply assumed the applicability of the Drude model, we have measured reflectance at multiple wavelengths over a region of (1-2.33 eV) and conclusively established a distinct free-electron behavior. That is, LMH reflectance increases as a function of increasing wavelength. The wavelength dependence observed is very well described with the Drude model (See fit in Fig.2 and Fig.S4).

Theoretically, the free electron model is applicable when the electron subsystem is both degenerate, that is T<< $T_f$ and the ratio of a dimensionless $\hbar/\varepsilon_f \tau$ is less than unity. The second criterion is related to the Boltzmann conduction transport theory where the scattering of electrons off ions is assumed to be a weak quantum perturbation to the system. In our experimental conditions, both of the above criteria are very well satisfied. Calculations of the micro-transport properties of LMH relevant to planetary conditions are provided below.

**Magnetic Reynolds number and magnetic diffusivity**

In magneto-hydrodynamics, the electrical conductivity $\sigma$ is usually expressed in terms of the magnetic diffusivity $\beta$ with $\beta = 1/\mu_0\sigma$, where $\mu_0$ is the magnetic vacuum permeability. The important quantity in dynamo action theory is the magnetic Reynolds number Rm, which assesses whether the dynamo action at a certain depth is possible. $R_m$ measures the ratio of magnetic field production to Ohmic dissipation and is equal $R_m = v_{conv}L/\beta$. Here, $v_{conv}$ is the convective flow velocity, $\beta$ is the magnetic diffusivity and L is the characteristic length scale for the conducting core. Dynamo numerical simulations suggest that $R_m$ must typically exceed 50 to guarantee dynamo action[42]. In the regime where the magnetic field is generated, $v_{conv}$ is predicted to be in the range of ~ 0.1 to 1 cm/s[7]. Since L is of the order of few $10^9$ cm, the largest acceptable value for $\beta$ is ~ $10^7$ cm$^2$/s for the outermost dynamo generating region. For the conductivities observed in our experiments, $\beta \sim 0.56 \times 10^4$ cm$^2$/s at pressures corresponding to 0.84 of Jupiter's radius and 0.64 of Saturn radius[28]. The addition of 10% helium is unimportant for these conductivity estimations, since at these P-T conditions helium still has a rather high ionization potential[29]. We note that these values are now very similar to recent estimates for the Earth iron core mantle boundary, yielding $\beta \sim 0.59 \times 10^4$ cm$^2$/s[43].

If metallization and dissociation occur continuously through the Jovian isentrope, then the depth corresponding $\beta = 10^7$ cm$^2$/s, is more likely shallower than previous estimates and may well extend to more than ~0.91$R_j$. This region is at roughly 7,000 km depth and corresponds to where the Jovian isentrope intersects the 50% dissociation line of hydrogen predicted by the DFT calculations[29]. We note that these values are larger than most previous estimates that place the source region of the Jovian dynamo at depths of 0.85-0.8$R_j$[14,16,44] and much larger than the usual definition of the Jovian metallic core[7].

**Thermal conductivity**

For degenerate simple metals at high temperatures, thermal conductivity, $\lambda$, is dominated by electronic transport. The relation between electrical and thermal conductivity in metals should be well described by the Wiedemann-Franz. Using the Lorenz number **L**$_0$, as proportionality constant $\lambda = \sigma L_0 T$, where T is the temperature in Kelvins. At conditions of the Jovian interior (4000K and 140-170 GPa), $\lambda \sim$ 100-140 W/mK. We note that there could be some systematic uncertainty due to the unknown dependence of $\sigma$ of LMH on temperature, though most theoretical treatments either assume or have shown some weak dependence. The ionic contribution to the thermal conductivity is known to be rather small and can be ignored. Our value is in very good agreement with recent theoretical calculations of $\lambda$ for LMH[19] but higher than those calculated along Jovian adiabat[29].

**Thermal diffusivity**

The important quantity that characterizes heat transport is thermal diffusivity, which is given by $\kappa = \lambda/\rho C_p$ where $C_p$ is the specific heat capacity at constant pressure and $\rho$ is the density in gm/cm$^3$. In the high temperature limit for liquid metals, $C_p$ can be assumed to be $3NK_B$ where N is the ion density at the condition of our experiment and $K_B$ is the Boltzmann constant. Like the conductivity, thermal diffusivity is dominated by electronic transport in a degenerate system, $\kappa \sim 1.3 \times 10^{-5}$ m$^2$/s